# A Model for Ball Lightning Derived from an Extension of the Electrodynamics Equations

Daniele Funaro[1]

1. *Dipartimento di Fisica, Informatica e Matematica, Università di Modena e Reggio Emilia, 41125 Modena, Italy*

**Forewords.** Most of the results here presented come from the extension of the equations of electrodynamics developed in [1], [2]. This is basically the coupling of the Maxwell's model with the equations ruling fluid motion. The new formulation strictly includes the classical one, allowing for an impressive enlargement of the space of solutions. Before starting, it is important to remark that we are in vacuum, without the presence of sources, such as for instance given currents or magnets. Thus, our aim is for the moment to describe pure EM radiations. The novelty is to suppose that the divergence of the electric field may get values different from zero. Failing to accept this assumption, brings back to the standard Maxwell's equations, making our extension useless. Thus, interesting situations will be examined with the requirement that div**E** is not vanishing.

**The model.** As usual **E** and **B** denote the electric and magnetic field, respectively. We set by definition: $\rho = \text{div}\mathbf{E}$. In addition, we have a new velocity field **V**, which has the role of indicating direction and speed of the energy flow (a sort of generalization of the Poynting vector). The system of equations is:

$$\frac{\partial \mathbf{E}}{\partial t} = c^2 \text{rot}\mathbf{B} - \rho \mathbf{V}$$

$$\frac{\partial \mathbf{B}}{\partial t} = -\text{rot}\mathbf{E}$$

$$\text{div}\mathbf{B} = 0$$

$$\rho\left(\frac{\partial \mathbf{V}}{\partial t} + (\mathbf{V}\cdot\nabla)\mathbf{V} + \mu(\mathbf{E} + \mathbf{V}\times\mathbf{B})\right) = -\nabla p$$

$$\frac{\partial p}{\partial t} = \mu\rho\, \mathbf{E}\cdot\mathbf{V}$$

where $c$ denotes the speed of light and, up to dimensional scaling, the scalar $p$ acts as a pressure. The constant $\mu$ has the dimension of charge divided by mass and has been estimated to be approximately $2.85\times 10^{11}$ Coulomb/Kg. A charge density can be defined by setting $\varepsilon_o\rho$, where $\varepsilon_o$ is the dielectric constant in vacuum. Formally, it is also possible to introduce a density of mass as $\varepsilon_o(\rho - p/c^2)/\mu$, although there is no real matter for the moment. This setting is fully consistent with all the relevant physical properties (see again [1], [2]). We recognize the Euler's equation with a forcing term given by the vector $\mathbf{E} + \mathbf{V} \times \mathbf{B}$, which recalls the Lorentz's law. By taking the divergence of the first equation, we get the continuity equation for the charge density, so that such a constraint does not need to be imposed independently. The last relation says that pressure may raise as a consequence of a lack of orthogonality between $\mathbf{E}$ and $\mathbf{V}$. The revised set of equations is similar to that ruling plasma physics (see e.g. [3], chap.10), with the difference that no effective matter is present in our case. We trivially return to the Maxwell's case in vacuum by imposing $\rho = 0$ and $p = 0$. The important fact here is that the vector field $\mathbf{V}$ is one of the unknowns and it is not directly related to the product $\mathbf{E} \times \mathbf{B}$, as suggested by the classical theory.

**Free waves.** A special case is obtained when $D\mathbf{V}/Dt=0$ and $p=0$. In this circumstance we have (recall that $\rho = \mathrm{div}\mathbf{E}$):

$$\frac{\partial \mathbf{E}}{\partial t} = c^2 \mathrm{rot}\mathbf{B} - \rho\mathbf{V}$$

$$\frac{\partial \mathbf{B}}{\partial t} = -\mathrm{rot}\mathbf{E}$$

$$\mathrm{div}\mathbf{B} = 0$$

$$\rho\,(\mathbf{E} + \mathbf{V} \times \mathbf{B}) = 0$$

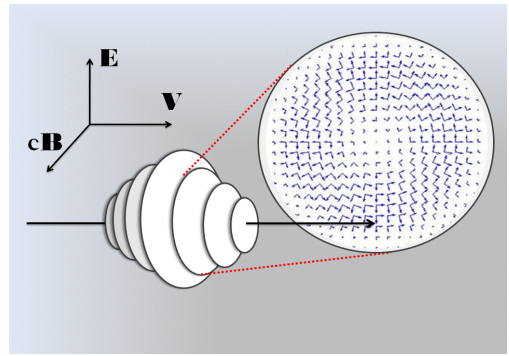

**FIGURE 1.** A special subset of solutions is obtained by requiring that $\mathbf{V}$ has no acceleration. Non-dissipative compact-support EM waves, travelling straightly at the speed of light, can be now modelled. Each travelling front displays orthogonal EM fields with div$\mathbf{B}$=0, however div$\mathbf{E}$ is not necessarily equal to zero.

This version (see Fig. 1) produces solutions called *free waves*. In this situation, it is possible that $\mathbf{V}$ is a gradient of a potential $\Psi$. Since the constancy of the speed of light leads to the *eikonal* equation: $|\mathbf{V}|=|\mathrm{grad}\Psi|=c$, *here* the waves exactly follow the rules of geometrical optics. The main consequence is the possibility of simulating 'photons' (i.e., non-diffusive EM emissions behaving like particles) in the framework of classical theories, reopening the discussion on the problem of wave-particle duality. Such solutions are not present in the standard Maxwell's theory ($\rho = 0$). It is shown that the integral (extended on the support of a photon) is zero, for both the density of charge and mass, in agreement with expectations.

**Vortex rings.** Since the new model equations are the results of the coupling with fluid equations, other more complicated structures may be taken into consideration. Among these, *vortex rings* have a special attention due to their extreme stability (see Fig. 2). In [1], [2], [4], solutions representing EM waves trapped in rings have been proposed. These are in general the mathematical combination of a dynamical component and a stationary one. The two pieces sum up linearly in simple situations, but they are mixed up in more serious cases. Each one cannot exist without the other. The dynamical component is somehow transparent, by meaning that the evolution of the corresponding EM fields displays zero average in space and time. The stationary component is needed for stability and provides the object with effective electric and magnetic properties. It displays a field **B**, whose lines are closed and turn around the main axis of the ring, and a radial electric field **E**. The stationary component, however, may contribute only in part to the whole energy. Again, here we are talking about pure EM waves (in the extended meaning of the term, according to the new model equations), since no charged massive particles have been introduced so far.

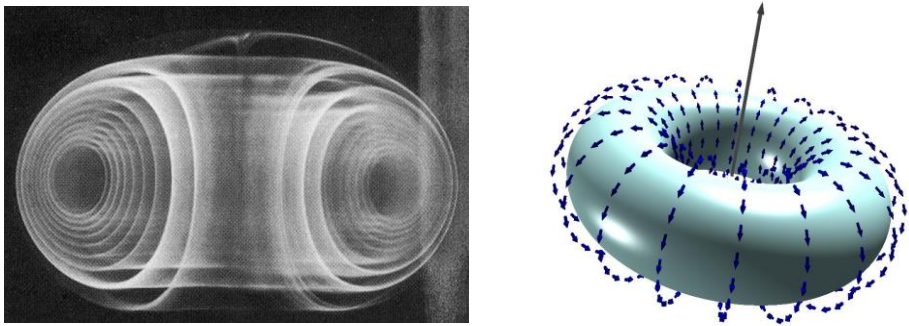

**FIGURE 2.** Fluid vortex ring colored with dye (courtesy of T.T. Lim, Nat. Univ. of Singapore). Similar behavior is obtained in the framework of EM waves with the help of the new set of equations.

Stability of EM ring structures may be deduced through very classical arguments. Each point of the rotating immaterial fluid, described by the velocity **V**, can be imagined to be an infinitesimal entity carrying density of charge different from zero. A magnetic field is present, so that the stream evolves along curbed lines due to the Lorentz's force. This force actually acts on the points, since they also carry a nonvanishing density of mass. An internal pressure is generated and the surface of the ring is characterized by having the gradient of pressure equal to zero.

Another way to explain stability is to argue in terms of general relativity. The model equations are obtained by energy and momentum preservation. In fact, they are recovered from the sum of the EM stress tensor with a suitable mass type tensor (see [1]). This sum can be put on the right-hand side of Einstein's equations in or-der to get the metric. It turns out from the theory that the scalar curvature $R$ of the metric is proportional to the pressure potential $p$. If it happens that the geodesics are the stream lines of the ring, then the structure should be stable. Thus, some photons are following the space-time geodesics created by their own motion. Such peculiar situations are not always obtained, but only when suitable conditions of energy and geometry are satisfied. Note that in the case we are discussing there are no classical masses at all, by the way the curvature of the space-time is a consequence of the presence of a nontrivial right-hand side in Einstein's equations.

We denote by $\eta$ the major diameter of the ring and by $q$ its global charge. Combining (105) and (111) in Appendix G in [2], one arrives at the following relation:

$$|q| \approx 42.35 \, \eta \, \varepsilon_o \, c^2/\mu \approx 1.18 \, \eta \, 10^{-4} \qquad (1)$$

For example, when $\eta$ is of the order of $10^{-15}$ meters, one finds that $|q|=e$ is approximately the charge of the electron. In this situation, by integrating the mass density on the volume one finds the global mass, which agrees with that of an electron. Of course, *spin* is directly related to the rotation of the dynamical part of the wave. This suggests the possibility of modelling electrons within a pure EM context. Toroid models of the electron are present in the literature (see [2], section 2.1, for a review). The proposal discussed here seems to fulfil all the basic features.

**Plasma dynamics.** We can now improve a bit our knowledge about plasmas in this new framework. We first introduce electrons and ions in the model. This is easily done through the standard equations (see [3]) and by suitably adding terms such as **V**div**E**. In this fashion we have the combined action of the density of charge carried by the particles (moving at speed **v**) and that of the radiation evolving at speed **V**. Indeed, it is not surprising that the electric field in the void interspace between charged particles may assume values of $\rho$ different from zero. This is a consequence of the finiteness of the speed of light, especially when high frequencies or fast movement of particles are involved. It is true that Coulomb's law guarantees div**E**=0 between particles, nevertheless it requires the information to travel at infinite speed. Therefore, in the new version, the Gauss theorem should be interpreted in milder form. The movement of massive particles is now fully coupled with the exchange of EM radiations. It is like dealing with three species (electrons, positive ions, pure waves). Of course, this extension is made in such a way that for div**E**=0 one returns to the usual plasma model.

**Ball lightning.** According to what has been said above, a pure EM vortex ring carries an intrinsic internal density of charge. Real charged particles (i.e., made of matter) can enter the toroid and find their equilibrium. They can move relatively slow, or they can even remain steady. Everything is ruled by the model that combines the standard plasma development with the exchange of EM information. As a consequence, the total charge comes from averaging the contribution of the actual physical charges with that naturally generated at the interior of the ring itself. The whole structure is stable for the reasons we specified before. For example, assuming that a negative density of charge is responsible for the stability of the ring, a certain number of positive ions can be stored into it, without reciprocal repulsion. We recall once again that we have a confined plasma into a container that has no physical boundaries. There is indeed an internal pressure that arrives with zero gradient at the border. These statements are waiting however to be confirmed by numerical experiments. If the object has a diameter $\eta \approx 10^{-1}$ meters (typical magnitude of a BL), we get from (1) that $q$ is of the order of $10^{-5} \div 10^{-4}$ Coulomb (in rough agreement with observations, see e.g. [5], [6]).

By computing the energy of the stationary component plus that of the ions, we get a quantity far below the level of what it is measured in BL. We have to remember however that a dynamical component (having $\rho = 0$, see [4]) also exists and may carry a large amount of energy. We guess that a BL is the addition of a very energetic EM rotating wave (with zero kinetic energy) with the stationary part mentioned above. The two combine in order to stay confined in a portion of space. Exact computations can be performed for rings having the minor diameter much smaller than the major one. In this case, the coupling between the stationary and dynamical parts is very mild, so that the last quantity can in principle reach arbitrary intensities. By impacting with the environment a BL can dissipate electromagnetic energy and diminish the magnitude of the dynamical rotating component. Radio waves are emitted during this process. Their frequency is inversely proportional to the diameter $\eta$ and are of the order of GHz when $\eta \approx 10^{-1}$ meters (the wave-length of such a cavity). When most of the energy is dissipated, the dynamical part of the trapped wave does not provide anymore a good support to the stationary part, so that the object becomes unstable. Without the negative constraining force, the positive ions tend to repel creating an explosion.

**Other comments.** Due to viscosity, fluid vortex rings tend to shift naturally in the direction of the axis. This is expected to be true also for plasma rings. Moreover, these structures are charged, so that they may follow specific (invisible) electric patterns in the atmosphere, especially if air conditions are favorable. Rings can take the shape of a ball (but showing a different topology, however) when their hole is reduced to a segment (Hill's vortex). In this situation, the study of the coupling of the different components (stationary and dynamical EM parts, plus ions) becomes rather involved. We have no conjectures at the moment about the reason why BL emit radiations in the visible range. Earlier models describing BL as EM bubbles were discussed for instance in [7], [8], but turned out to be not fully satisfactory. In our opinion, if we want to follow this direction of research in BL, the extension of the EM equations is a necessary step to reach more convincing results.